\documentclass[conference]{IEEEtran}

\usepackage{cite}
\usepackage{amsmath,amssymb,amsfonts}
\usepackage{amssymb}
\usepackage{algpseudocode}
\usepackage{amsfonts}
\usepackage{graphicx}
\usepackage{fancyhdr}
\usepackage{cases}
\usepackage{textcomp}
\usepackage{extarrows}
\usepackage{algorithm,algpseudocode}
\usepackage{multirow}
\usepackage{graphicx}
\usepackage{overpic}
\usepackage{graphicx}
\usepackage{booktabs}
\usepackage[caption=false,font=footnotesize]{subfig}
\IEEEoverridecommandlockouts

\usepackage{orcidlink}
\hypersetup{hidelinks} 
\usepackage{multirow,tabularx}
\usepackage{mathtools}
\usepackage{xcolor}
\usepackage[english]{babel}
\usepackage{amsthm}

\IEEEoverridecommandlockouts
\def\BibTeX{{\rm B\kern-.05em{\sc i\kern-.025em b}\kern-.08em
    T\kern-.1667em\lower.7ex\hbox{E}\kern-.125emX}}
    
\begin{document}

\bstctlcite{BSTcontrol}
\title{One-Step Generative Channel Estimation  via Average Velocity Field}

\author{
    \IEEEauthorblockN{Zehua Jiang$^{1}$, Fenghao Zhu$^{1}$, Siming Jiang$^{1,2}$, Chongwen Huang$^{1}$, Zhaohui Yang$^{1}$, \\ Richeng Jin$^{1}$, Zhaoyang Zhang$^{1}$, and M\'{e}rouane~Debbah$^{3}$,~\IEEEmembership{Fellow,~IEEE}
    }
    \IEEEauthorblockA{$^1$ College of Information Science and Electronic Engineering, Zhejiang University, Hangzhou 310027, China}
    \IEEEauthorblockA{$^2$ Guangdong Tobacco Maoming Co., Ltd., Maoming, 525000, China}
    \IEEEauthorblockA{$^3$ KU 6G Research Center, Khalifa University of Science and Technology, P O Box 127788, Abu Dhabi, UAE}
    }
\maketitle

\begin{abstract}
Generative models have shown immense potential for wireless communication by learning complex channel data distributions. However, the iterative denoising process associated with these models imposes a significant challenge in latency-sensitive wireless communication scenarios, particularly in channel estimation. To address this challenge, we propose a novel solution for one-step generative channel estimation. Our approach bypasses the time-consuming iterative steps of conventional models by directly learning the average velocity field. Through extensive simulations, we validate the effectiveness of our proposed method over existing state-of-the-art diffusion-based approach. Specifically, our scheme achieves a normalized mean squared error up to 2.65 dB lower than the diffusion method and reduces latency by around 90\%, demonstrating the potential of our method to enhance channel estimation performance.

\end{abstract}


\section{Introduction}\label{sec:intro}
The six-generation (6G) wireless networks require high data rates, massive connectivity and low latency. Therefore, massive multiple-input multiple-output (MIMO) has been proposed to meet this vision \cite{wang2023road,zhu2025wireless}. However, its reliance on a large number of antennas introduces critical challenges, such as the demand for accurate and real-time channel state information (CSI). Specifically, this problem is exacerbated in mobile scenarios where rapid channel variations lead to the rapid obsolescence of CSI, severely degrading system performance \cite{wang2022time}. Consequently, the need for channel estimation methods with low latency has become the bottleneck of 6G wireless communication systems.
\par
To address this challenge, the research paradigm has shifted from conventional model-driven methods towards more powerful data-driven approaches \cite{arvinte2022score, balevi2020high}. Unlike model-driven methods that often rely on simplified assumptions like channel sparsity, data-driven models can learn the complex, underlying probability distribution of wireless channels from data. Among these, generative models such as generative adversarial networks (GANs) and diffusion models (DMs) have shown remarkable potential \cite{ho2020denoising,goodfellow2020generative}. For instance, GANs have been successfully employed to learn the channel manifold, mapping a low-dimensional latent space to the high-dimensional channel space \cite{zhang2022effective, wgan-gp}. More recently, score-based models like DMs which were originally applied in image generation have shown high accuracy in channel estimation \cite{dhariwal2021diffusion,ma2024diffusion}. Building upon this foundation, flow-based generative models were developed, reframing the denoising problem as learning the marginal vector field that defines a continuous path from the prior channel distribution to the target channel distribution \cite{lipman2023flow}. Fundamentally, both score-based approaches and flow-based models learn to utilize the prior distribution for posterior sampling, achieving robust performance even in out-of-distribution scenarios. Overall, generative approaches have shown huge potential in terms of estimation accuracy and robustness \cite{fan2025generative}.
\par
However, these advanced generative techniques share a common significant limitation: high latency caused by an iterative generative process \cite{song2020denoising}. For instance, GAN-based channel estimation iterates to find the optimal latent vector that corresponds to the observed channel \cite{du2023joint}. Similarly, DMs generate samples through a reverse multi-step denoising procedure, often involving computationally expensive numerical solvers for stochastic differential equations or Langevin dynamics \cite{song2020score,arvinte2022mimo}. This multi-step nature introduces high inference latency, which is a critical bottleneck for the real-time requirements of wireless communication systems. To solve this fundamental issue, we introduce a novel one-step generative framework for channel estimation based on flows \cite{geng2025mean}. We consider the marginal vector field as an instantaneous velocity field (IVF). Instead of iterating with the IVF, we propose to directly learn the average velocity field (AVF). The AVF represents a direct mapping from the prior channel distribution to the target channel distribution, enabling a high-fidelity channel estimation in a single number of function evaluations (NFE). This approach fundamentally bypasses the need for iterative sampling, drastically reducing latency while retaining the powerful distribution learning capabilities of generative models.

\par
In this work, the main contributions are summarized as follows:
\begin{itemize}
    \item We introduce a novel channel estimation algorithm which operates in a single feedforward step. This one-step framework fundamentally reduces inference latency by around 90\% at a signal-to-noise ratio (SNR) of 10 dB compared with iterative methods like diffusion models.
    \item The proposed approach does not require prior knowledge of SNR during inference, showing the robustness of our method across diverse channel conditions. 
    \item Evaluation results demonstrate that the AVF-based method outperforms the established baseline estimators. In particular, the proposed algorithm achieves an average 1.75 dB improvement in estimation accuracy compared with the state-of-the-art (SOTA) diffusion method. Especially in high SNR scenario, this improvement arrives at most 2.65 dB.
 
\end{itemize}
\par
The rest of this paper is organized as follows: Section \ref{sec:sys} describes the system model and problem formulation. Section \ref{sec:method} provides a comprehensive analysis of the one-step channel estimation method, including the concept of AVF and the training strategy of the proposed network. Simulation results are presented in Section \ref{sec:simulation} to verify the performance of the proposed algorithm. Finally, the conclusions are outlined in Section \ref{sec:conclusion}.

\vspace{4mm}

\section{System Model and Problem Formulation}\label{sec:sys}
Consider a MIMO system with an $N_{tx}$-antenna mobile terminal and an $N_{rx}$-antenna base station. We denote the unitary pilot matrix as $\mathbf{P} \in \mathbb{C}^{N_{\mathrm{tx}} \times N_{\mathrm{p}}}$, which satisfies $\mathbf{P}\mathbf{P}^\mathrm{H} = \mathbf{I}$. Regarding the channel matrix $\mathbf{H} \in \mathbb{C}^{N_{\mathrm{rx}} \times N_{\mathrm{tx}}}$, we consider $\mathbb{E}[\mathbf{H}]=\mathbf{0}$ and $\mathbb{E}[||\mathbf{H}||^2_2]=\mathrm{N}$. Therefore, the transmission model can be formulated as:
\begin{equation}
    \mathbf{Y}=\mathbf{H} \mathbf{P}+\mathbf{N} \in \mathbb{C}^{N_{\mathrm{rx}} \times N_{\mathrm{p}}},
    \label{eq:transmission}
\end{equation}
where $\mathbf{N} \sim \mathcal{N}_{\mathbb{C}}(\mathbf{0},\sigma^2 \mathbf{I})$ is additive white Gaussian noise (AWGN), a standard benchmark approximating aggregate real-world noise via the central limit theorem. Then we left-multiply equation (\ref{eq:transmission}) by $\mathbf{P}^\mathrm{H}$ to obtain the noisy channel observation
\begin{equation}
    \hat{\mathbf{H}}=\mathbf{Y} \mathbf{P}^{\mathrm{H}}=\mathbf{H}+\hat{\mathbf{N}},
\end{equation}
where $\hat{\mathbf{N}}=\mathbf{N} \mathbf{P}^{\mathrm{H}}$ remains AWGN with variance $\sigma^2$ due to $\mathbf{P}$ being unitary. Further, the observed channel $\hat{\mathbf{H}}$ and the ground truth channel $\mathbf{H}$ are transformed into angular domain using the Fast Fourier Transform (FFT)
\begin{equation}
\mathbf{H}_{\mathrm{ang}} ={\mathrm{FFT}}( \mathbf{H}),
\end{equation}
\begin{equation}
    \hat{\mathbf{H}}_{\mathrm{ang}} = \mathrm{FFT}(\hat{\mathbf{H}}).
\end{equation}
The network is trained using $\hat{\mathbf{H}}_{\mathrm{ang}}$ as initial input. This FFT transformation effectively converts a dense estimation problem into a sparse one. Notably, our approach is agnostic to the specific sparsity level and structure of the channel distribution, ensuring its broad applicability to various propagation scenarios.
\par
We assume $\mathrm{Model}(\cdot)$ to be any channel estimation methods, the estimated channel in angular domain can be written as 
\begin{equation}
    \hat{\mathbf{H}}_\mathrm{est} = \mathrm{Model} (\hat{\mathbf{H}}_{\mathrm{ang}}).
\end{equation}
We transform the angular estimation channel back into the spatial domain via the inverse FFT as
\begin{equation}
    \mathbf{H}_{\mathrm{est}}=\mathrm{IFFT}(\hat{\mathbf{H}}_{\mathrm{est}}).
\end{equation}
\par
For notational simplicity, the channel matrix $\mathbf{H}$ is assumed to be in the angular domain throughout the following section except Algorithm \ref{alg:Training} and Algorithm \ref{alg:Inference}.
\section{AVF-based Channel Estimation}\label{sec:method}

\subsection{Average Velocity Field}

\begin{figure}[t]\vspace{0mm}
    \centering
    \includegraphics[width=0.7\linewidth]{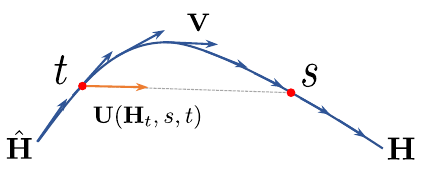}
    \vspace{-0mm}
    \captionsetup{name={Fig.}, labelsep=period}
    \caption{Channel estimation with AVF. The curved path shows the wireless flow trajectory, where the blue arrow is IVF \textbf{V}, and the orange arrow is the average velocity field.}
    \label{fig:AVF illustration}
    \vspace{-4 mm}
\end{figure}

Based on flows, we define a time-continuous channel transformation path that connects the ground-truth (clean) channel to the observed (noisy) channel. Formally, given time sequence $t \in [0,1]$, we denote $\mathbf{H}_t$ as an intermediate time-state on this path, a state transform equation can be written as
\begin{equation}
    \mathbf{H}_{t} = a_{t}\mathbf{H} + b_{t}\hat{\mathbf{H}},
    \label{eq: statetransform function}
\end{equation}
where $\mathbf{H}=\mathbf{H}_0$ and $\hat{\mathbf{H}}=\mathbf{H}_1$ denote the ground-truth channel and the observed channel respectively, $a_t$ and $b_t$ are predefined scalar schedules. The corresponding IVF of this path is the time derivative of $\mathbf{H}_t$
\begin{equation}
    \mathbf{V}_t = \frac{d}{dt}\mathbf{H}_t=\frac{da_t}{dt}\mathbf{H}+\frac{db_t}{dt}\hat{\mathbf{H}}.
\end{equation}
A common choice for the schedules is $a_t=1-t$ and $b_t=t$, which simplifies the IVF to $\mathbf{V}_t=\hat{\mathbf{H}}-\mathbf{H}$. In general, this process is governed by the ordinary differential equation (ODE)
\begin{equation}
    \frac{d}{dt}\mathbf{H}_{t} = \mathbf{V}(\mathbf{H}_{t}, t).
\end{equation}
\par
The IVF at a given time-state $\mathbf{H}_t$, denoted by $\mathbf{V}(\mathbf{H}_t, t)$, represents the tangent denoising direction, i.e., field element $v_{ij}$ represents the tangent denoising direction of channel element $(h_t)_{ij}$. This is the quantity that conventional flow-based models aim to learn. The ground-truth channel can then be recovered by subtracting the integral of IVF from $t=0$ up to $t=1$
\begin{equation}
    \mathbf{H}_0 = \mathbf{H}_1 - \int_{0}^{1} \mathbf{V}(\mathbf{H}_{\tau}, \tau) d\tau.
\end{equation}
However, numerically approximating this integral typically requires a multi-step solver, which leads to high latency.
\par
To enable one-step estimation, we introduce the concept of AVF, denoted by $\mathbf{U}(\mathbf{H}_{t}, s, t)$. As illustrated in Fig. \ref{fig:AVF illustration}, this field represents the average of the integrated instantaneous velocities over the time interval from $s$ to $t$
\begin{equation}
    \mathbf{U}(\mathbf{H}_{t}, s, t) \triangleq \frac{1}{t-s} \int_{s}^{t} \mathbf{V}(\mathbf{H}_{\tau}, \tau) d\tau.
    \label{eq:avg_velocity}
\end{equation}
By multiplying $(t-s)$ on both sides of equation (\ref{eq:avg_velocity}) and subsequently differentiating with respect to $t$, we obtain
\begin{equation}
    \frac{d}{d t}\left[(t-s) \mathbf{U} (\mathbf{H}_t, s, t )\right]=\frac{d}{d t} \int_s^t \mathbf{V}  (\mathbf{H}_\tau, \tau ) d \tau.
    \label{eq: derivative}
\end{equation}
Applying the product rule to the left-hand side and the fundamental theorem of calculus to the right-hand side of equation (\ref{eq: derivative}) and rearrange the derivative term, while treating $s$ as independent of $t$, we arrive at the central function for the AVF
\begin{equation}
    \mathbf{U} (\mathbf{H}_t, s, t )=\mathbf{V} (\mathbf{H}_t, t )-(t-s)\frac{d}{d t} \mathbf{U} (\mathbf{H}_t, s, t ).
    \label{AVF}
\end{equation}
The primary problem is now to compute the total derivative term $d\mathbf{U}(\cdot)/dt$ in equation (\ref{AVF}). This total derivative can be expanded in terms of its partial derivatives using the chain rule
\begin{equation}
    \frac{d}{d t} \mathbf{U} (\mathbf{H}_t, s, t )=\frac{d \mathbf{H}_t}{d t} \partial_{\mathbf{H}} \mathbf{U}+\frac{d s}{d t} \partial_s \mathbf{U}+\frac{d t}{d t} \partial_t \mathbf{U}.
    \label{eq:partial derivatives}
\end{equation}
It is noted that $d\mathbf{H}_t / dt = \mathbf{V}(\mathbf{H}_t,t)$, $ds/dt=0$, and $dt/dt=1$. Therefore, we can simplify equation (\ref{eq:partial derivatives}) into
\begin{equation}
    \frac{d}{d t} \mathbf{U} (\mathbf{H}_t, s, t )=\mathbf{V}(\mathbf{H}_t,t) \partial_{\mathbf{H}} \mathbf{U}+\partial_t \mathbf{U}.
    \label{eq:simplified partial derivatives}
\end{equation}
Equation (\ref{eq:simplified partial derivatives}) demonstrates that the total derivative can be computed via the Jacobian-vector product (JVP). This operation involves the product of the Jacobian matrix of the function $\mathbf{U}$, denoted as $[\partial_{\mathbf{H}} \mathbf{U}, \partial_s \mathbf{U}, \partial_t \mathbf{U}]$, and the corresponding tangent vector $[\mathbf{V}, 0, 1]^\top$.
\subsection{AVF for Network Training and Inference}
With the theoretical framework established, we now formulate the training objective. We employ a convolutional neural network (CNN) with parameters $\theta$ to learn $\mathbf{U}_\theta$. The model is trained by minimizing a loss function that regresses the output  of network onto a target derived from the AVF property. The loss function is defined using the MSEloss
\begin{equation}
    \mathcal{L}(\theta) = \mathrm{MSE}(\mathbf{U}_{\theta}(\mathbf{H}_{t}, s, t),\mathrm{sg}(\mathbf{U}_{\mathrm{target}})),
\end{equation}
where the regression target, $\mathbf{U}_{\mathrm{target}}$, is constructed to satisfy the AVF relationship from equation (\ref{AVF})
\begin{equation}
    \mathbf{U}_{\mathrm{target}} = \mathbf{V} (\mathbf{H}_t, t )-(t-s) (\mathbf{V}(\mathbf{H}_t,t) \partial_{\mathbf{H}} \mathbf{U}_{\theta} + \partial_t \mathbf{U}_{\theta} ).
\end{equation}
Notably, to maintain a computationally tractable training process, we apply a stop-gradient operator denoted as $\mathrm{sg}(\cdot)$ to the target term. This operation avoids the computational overhead for a double back-propagation through the Jacobian-vector product.

\addtolength{\topmargin}{+.10in}

\begin{algorithm}[t]
\caption{AVF: Training}
\label{alg:Training}
\begin{algorithmic}[1]
    \State \textbf{Input:} Time slot: $\mathrm{T}$, ground truth channel: $\mathbf{H}$, snr list: $\mathrm{SNR}$.
    \State Initialize the $\mathrm{CNN}_{\theta}$ model weights $\theta$.
    \For{$i\leftarrow 1,2,\cdots,n$}
        \State $\mathrm{snr} = \mathrm{sample}(\mathrm{SNR}) $
        \State $s,t = \mathrm{sample\_time\_steps()}$
        \State $\hat{\mathbf{N}} = \mathbf{H} / \sqrt{\mathrm{snr}}$
        \State $\hat{\mathbf{H}}=\mathbf{H}+\hat{\mathbf{N}}$
        \State $\mathbf{H}_{\mathrm{ang}}={\mathrm{FFT}} ( \mathbf{H} )$
        \State $\hat{\mathbf{H}}_{\mathrm{ang}} = \mathrm{FFT}(\hat{\mathbf{H}})$
        \State $\mathbf{H}_t = (1-t) \times \mathbf{H}_{\mathrm{ang}} + t \times \hat{\mathbf{H}}_{\mathrm{ang}}$
        \State $\mathbf{V} = \hat{\mathbf{H}}_{\mathrm{ang}} -\mathbf{H}_{\mathrm{ang}}$
        \State $(\mathbf{U}_{\theta}, d\mathbf{U}/dt)=\mathrm{JVP}(\mathrm{CNN},(\mathbf{H}_t, s,t), (\mathbf{V},0,1))$
        \State $\mathbf{U}_{target}=\mathbf{V}-(t-s)\times d\mathbf{U}/dt$
        \State $\mathcal{L}(\theta) = \mathrm{MSE}(\mathbf{U}_{\theta}(\mathbf{H}_{t}, s, t),\mathrm{sg}(\mathbf{U}_{\mathrm{target}}))$
        \State Update $\theta$ as (\ref{Adam})
    \EndFor
    \State \Return $\theta^*$
    
\end{algorithmic}
\end{algorithm}

\begin{algorithm}[t]
\caption{AVF: Inference}
\label{alg:Inference}
\begin{algorithmic}[1]
    \State \textbf{Input:} $\mathbf{Y}$, $\mathrm{NFE}$.
    \State Load the $\mathrm{CNN}_{\theta}$ model weights $\theta^*$
    \State $\hat{\mathbf{H}}_{\mathrm{init}}=\mathbf{Y} \mathbf{P}^{\mathrm{H}}$
    \State $\hat{\mathbf{H}}_{\mathrm {ang}}=\operatorname{\mathrm{FFT}}(\hat{\mathbf{H}}_{\mathrm{init}})$
    \For{$t \leftarrow \mathrm{NFE},\cdots,2,1$}
    \State $\hat{\mathbf{H}}_\mathrm{est} = \hat{\mathbf{H}}_{ang} - \mathrm{CNN}_{\theta^*} (\hat{\mathbf{H}}_{ang}, \frac{t-1}{\mathrm{NFE}}, \frac{t}{\mathrm{NFE}})$
    \EndFor
    \State $\mathbf{H}=\mathrm{IFFT}(\hat{\mathbf{H}}_\mathrm{est})$
    \State \Return $\mathbf{H}$
    \label{Al:inference}
    
\end{algorithmic}
\end{algorithm}
\vspace{-0 mm}

The network parameters $\theta$ are updated iteratively using the Adam optimizer. The update rule is given by
\begin{equation}
    \theta^{*} = \theta + \alpha \cdot \mathrm{Adam}(\nabla_{\theta} \overline{\mathcal{L}}, \theta),
    \label{Adam}
\end{equation}
where $\alpha$ is the learning rate, $\overline{\mathcal{L}}$ represents the loss averaged over a single batch, $\theta^{*}$ is the updated parameters.
\par

During training, time step pairs $(s, t)$ are sampled from a logit-normal distribution parameterized by $\mu_t$ and $\sigma_t$. We introduce a mixing probability $p = (s \neq t)$ to balance the learning of the average velocity $\mathbf{U}$ and the instantaneous velocity $\mathbf{V}$ (corresponding to $s=t$). Notably, setting $p=0$ reduces the framework to conventional flow matching, while a non-zero $p$ activates the AVF mechanism. Additionally, randomizing the SNR for each batch ensures model robustness across diverse noise conditions. The overall procedure is summarized in Algorithm \ref{alg:Training}.
\par
Once trained, the AVF model enables an efficient, one-step inference. Estimation of the channel at any time-state $s$ from another time state $t$ can be performed directly. Given an input channel state $\mathbf{H}_{t}$, the model estimate the target channel state $\mathbf{H}_{s}$ via the predicted $\mathbf{U}_{\theta^*}$
\begin{equation}
    \mathbf{H}_{s} = \mathbf{H}_{t} - (t-s)\mathbf{U}_{\theta^*}(\mathbf{H}_{t}, s, t).
\end{equation}
For the 1-NFE case, we simply set $s=0$ and $t=1$. Besides, we also note that the proposed algorithm natively supports multiple NFE by equally dividing the entire path into multiple sub-paths. Algorithm \ref{alg:Inference} summarizes the inference algorithm.
\section{Simulation results}\label{sec:simulation}
In this section, we evaluate the performance of the proposed method in comparison to baseline approaches. We first illustrate the detailed system settings, followed by the presentation of the simulation results.
\subsection{System Configs}
Our system configs are shown as Table \ref{tab:system configs}, where $\alpha$ refers to the learning rate. Although we train total epochs of 8,000, model weights are saved periodically. For testing, we selected the checkpoint corresponding to the loss convergence point at approximately 1000 epochs. To ensure a fair comparison, we adopt the same backbone CNN architecture as the diffusion baseline \cite{fesl2024diffusion} with 55k parameters. In addition, the dataset is generated from the 3rd Generation Partnership Project (3GPP) spatial channel model \cite{3gpp}. To improve performance, we apply an exponential moving average to the model parameters. This operation smooths the weights against optimization noise, leading to a more robust final model.

\begin{table}[t]\vspace{0mm}
\centering
\caption{Configurations}\vspace{-0mm}
\begin{tabular}{c cc c} 
\toprule
configs & Value & configs & Value\\ [0.2ex] 
\midrule
$N_{\text{rx}}$ & 16 & SNR max (dB) & 30 \\
$N_{\text{tx}}$ & 64 & SNR min (dB) & -10 \\
$\mu_t$ & 0.4 & SNR steps (dB) & 5\\
$\sigma_t$ &  1.0 & epochs & 8000\\
$\alpha$ & 1e-4 & batch size & 512 \\
$p$ &  25\%  & warmup steps &  500   \\
\bottomrule
\end{tabular}\vspace{0mm}
\label{tab:system configs}
\end{table}

\begin{table}[t]
\vspace{0mm}
\centering
\caption{Complexity for $(N_{rx}, N_{tx}) = (64, 16)$.}\vspace{-0mm}
\begin{tabular}{c c c} 
\toprule
Method & Parameters & Online Complexity \\ [0.2ex] 
\midrule
LMMSE  & $1.05\cdot 10^6$ &  $\mathcal{O}(N^2)$\\ 
GMM \cite{fesl2022channel}  & $1.35 \cdot 10^8$  & $\mathcal{O}(KN^2)$  \\  
Diffusion \cite{fesl2024diffusion} & $5.50 \boldsymbol{\cdot} 10^{4}$ & $\mathcal{O}(N(\hat{l} k^2 C_{\mathrm{max}}^2 + \mathrm{log}N))$  \\
AVF & $ 5.50 \boldsymbol{\cdot} 10^{4}$ & $\mathcal{O}(N(k^2 C_{\mathrm{max}}^2 + \mathrm{log}N))$  \\
\bottomrule
\end{tabular}
\vspace{-4mm}
\label{tab:complexity}
\end{table}

We perform training and inference on a computer equipped with Ubuntu 22.04, an EPYC 9654 CPU, and NVIDIA A100 GPUs, using PyTorch 2.6 and Python 3.10. To serve as baselines, other algorithms are listed below.
\begin{itemize}
    \item \textbf{Baseline 1} (LS): The Least Squares (LS) is the most fundamental channel estimator which requires no prior knowledge of the channel statistics.
    \item \textbf{Baseline 2} (LMMSE): The Linear Minimum Mean Square Error (LMMSE) improves upon the LS method by utilizing the prior knowledge of the channel, specifically the channel covariance matrix estimated from the training data.
    \item \textbf{Baseline 3} (GMM): The Gaussian Mixture Model (GMM) is a probabilistic model with Kronecker product covariance \cite{fesl2022channel}, which represents a collection of data points as a mixture of several Gaussian distributions.
    \item \textbf{Baseline 4} (Diffusion): We include a SOTA generative channel estimator based on diffusion \cite{fesl2024diffusion}, which performs through a reverse denoising iterative process.
\end{itemize}
\par
Table \ref{tab:complexity} compares the computational complexity and parameter counts of various methods. For CNN-based approaches, the complexity comprises the network inference term $\mathcal{O}(N k^2 C_{\mathrm{max}}^2)$ and the FFT overhead $\mathcal{O}(N \log N)$, where $N=N_{rx}N_{tx}$, $k$ is the kernel size, and $C_{\max}$ is the maximum channel size. A critical distinction lies in the inference steps: while the diffusion baseline requires $\hat{l}$ iterative steps \cite{fesl2024diffusion}, the proposed AVF method achieves single-step inference, thereby significantly reducing computational cost. For the GMM baseline, complexity scales with the number of Gaussian components $K$ \cite{fesl2022channel}.
\subsection{Simulation Results}
\begin{figure}[t]\vspace{0mm}
    \centering
    \includegraphics[width=0.97\linewidth]{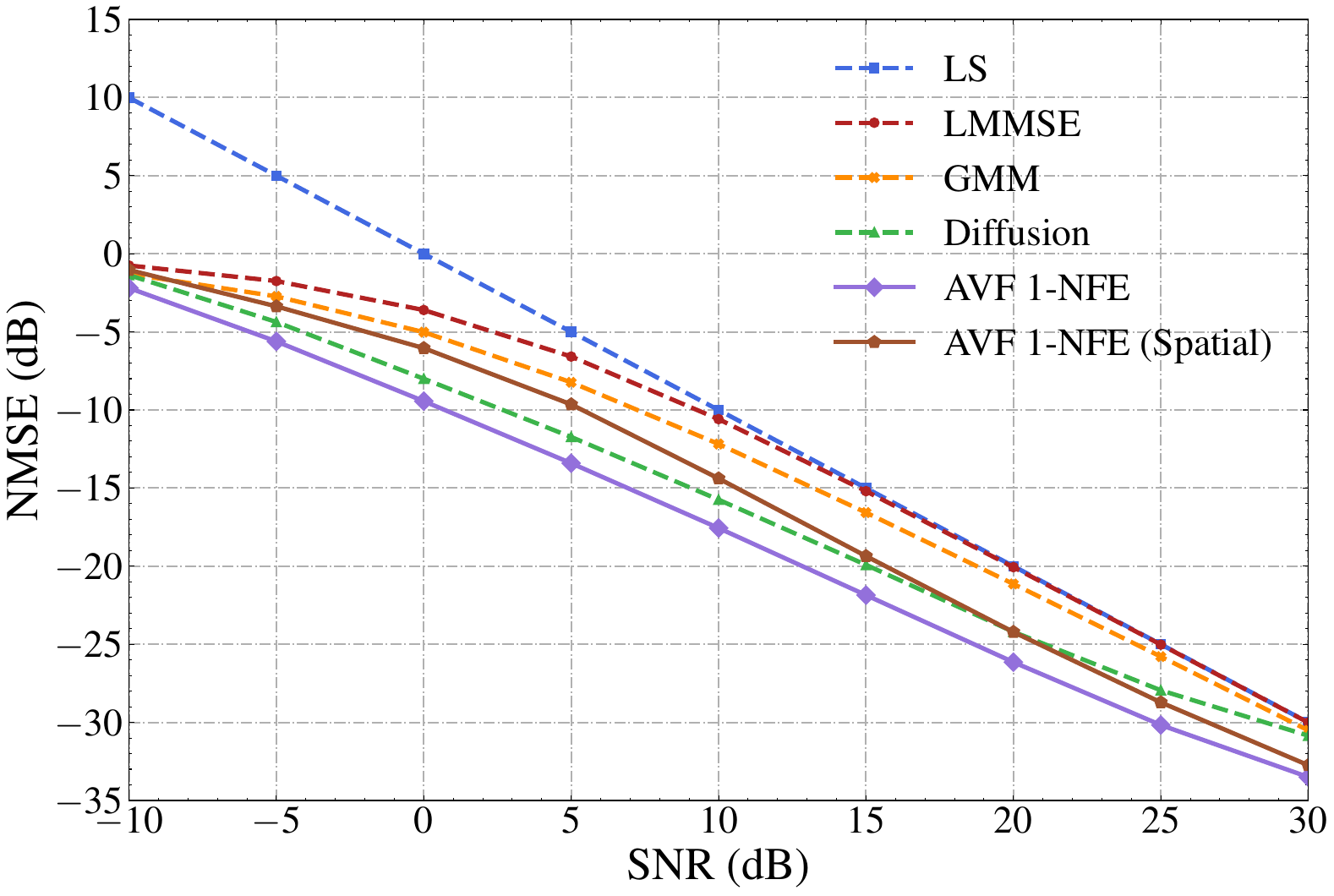}
    \captionsetup{name={Fig.}, labelsep=period}
    \caption{Performance evaluation with respect to SNR.}
    \label{fig:AVF}
    \vspace{-0 mm}
\end{figure}

\begin{figure}[t]\vspace{0mm}
    \centering
    \includegraphics[width=0.98\linewidth]{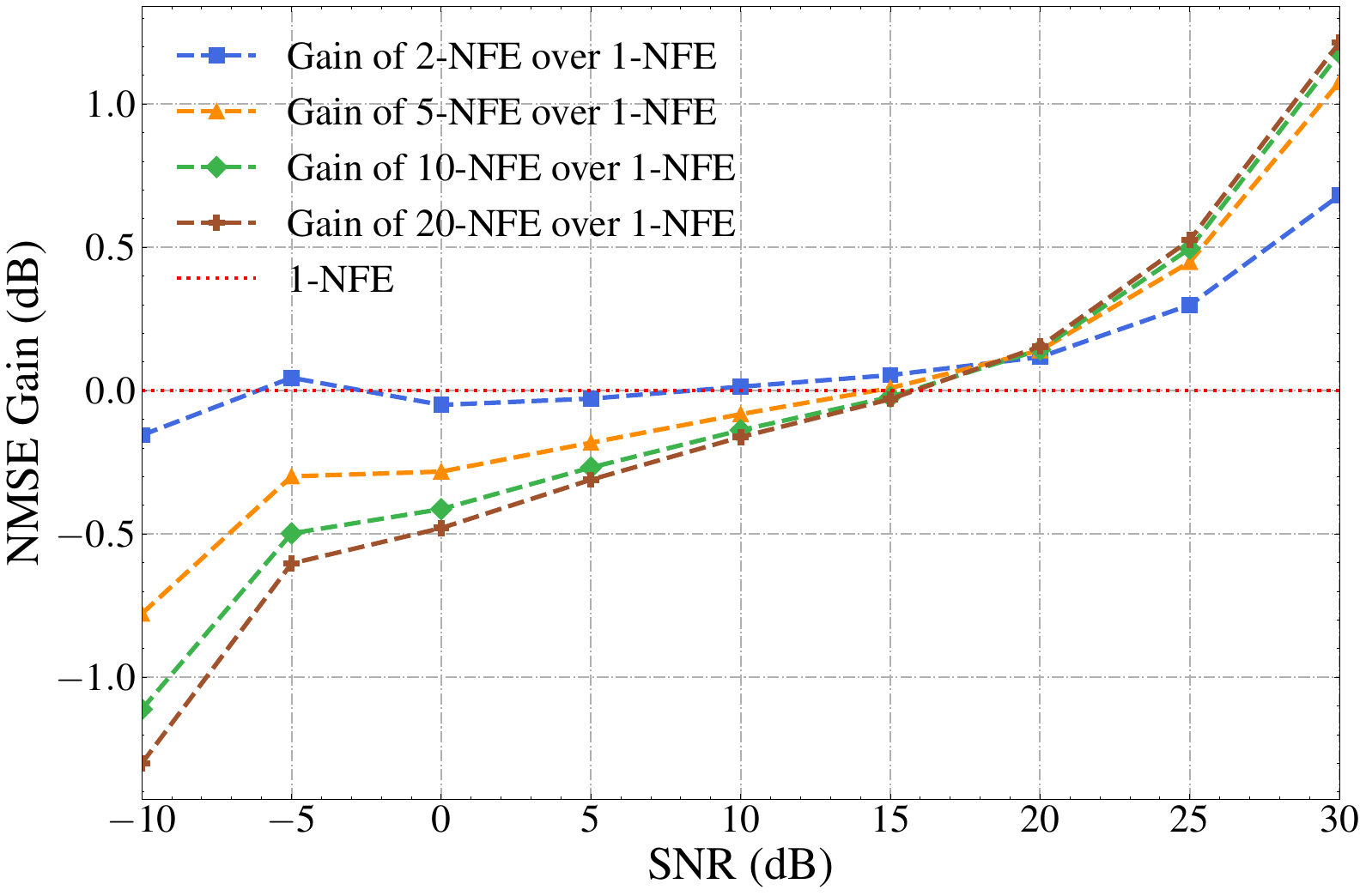}
    \captionsetup{name={Fig.}, labelsep=period}
    \caption{NMSE Gain of different NFE over 1-NFE.}
    \label{fig:NFE}
    \vspace{-4 mm}
\end{figure}

\begin{figure}[t]\vspace{0mm}
    \centering
    \includegraphics[width=0.98\linewidth]{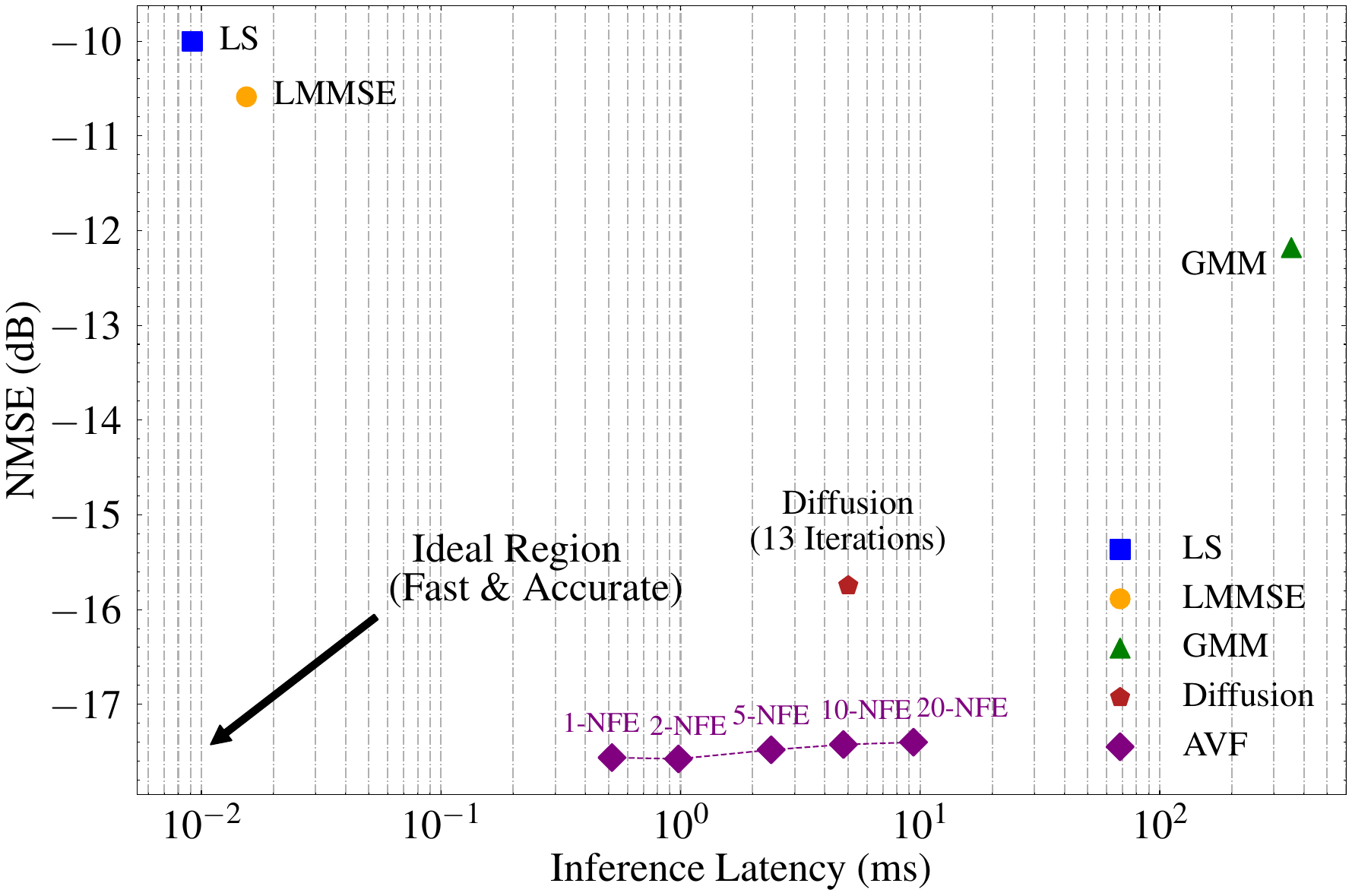}
    \captionsetup{name={Fig.}, labelsep=period}
    \caption{NMSE and inference latency of various methods when SNR = 10 dB.}
    \label{fig:Latency}
    \vspace{-4 mm}
\end{figure}
Fig. \ref{fig:AVF} evaluates the NMSE performance of different methods. It is evident that the AVF method with 1-NFE outperforms all estimators across the entire SNR region. Compared to the diffusion method, the proposed AVF method consistently achieves an NMSE improvement of over 1 dB across the entire tested SNR range, with the gain reaching 2.65 dB at an SNR of 30 dB. Notably, this performance gap widens as the SNR increases, indicating that our method more effectively leverages higher signal quality for enhanced estimation accuracy. We also evaluate the GMM estimator with an optimal configuration from \cite{fesl2022channel}. It is obvious that while the GMM estimator achieves good performance at an SNR of -10 dB, its performance degrades significantly as the SNR becomes larger. Furthermore, the AVF spatial counterpart achieves relatively low accuracy, showing the superiority of transforming the channel into angular domain, which suggests that exploiting sparsity of the channel in the angular domain is highly beneficial. 
\par
Fig. \ref{fig:NFE} illustrates the NMSE gain of multiple NFE strategies over the one-step baseline, revealing divergent trends across SNR regimes. In low SNR conditions, increasing NFE degrades performance, evidenced by a 1.3 dB loss at -10 dB when using 20-NFE. 
Conversely, high SNR regions benefit from multi-step evaluation, where a 2-NFE configuration at 30 dB achieves a 0.68 dB gain that increases with additional steps while showing diminishing marginal returns suggesting convergence. This contrast highlights a trade-off between minimizing numerical discretization errors and mitigating accumulated model prediction errors. Specifically, the high uncertainty of predicted velocity at low SNR causes iterative integration to accumulate errors and drift, whereas the 1-NFE approach effectively filters these fluctuations via the global average velocity. In contrast, the precise velocity field learned at high SNR allows multi-step refinement to dominate by reducing discretization error, thereby enhancing estimation accuracy.
\par
Fig. \ref{fig:Latency} presents the accuracy and latency of different algorithms processing a single channel sample at $\text{SNR}=10$ dB. The results highlight that traditional methods like LS and LMMSE, while extremely fast, exhibit poor NMSE performance. Besides, the GMM-based approach is computationally prohibitive, requiring 354.31 ms while only achieving a moderate NMSE of -12.18 dB. As a strong generative baseline, the diffusion model achieves a -15.75 dB performance level using an average of 13 iteration steps (5.01 ms). Crucially, this multi-step process is intrinsic to diffusion models, which require fine-grained steps to accurately approximate the reverse dynamics. Reducing the NFEs to match the low latency would lead to severe performance degradation. In contrast, the proposed AVF method demonstrates superior performance by learning a direct straight trajectory. Specifically, with just a single function evaluation, it achieves a low NMSE of -17.56 dB in only 0.52 ms. This makes the AVF estimator reduce computation time by nearly 90\% compared to the diffusion model, while providing a gain of 1.8 dB.
\section{Conclusion}\label{sec:conclusion}
In this paper, an one-step channel estimation method leveraging AVF was proposed. This approach requires no prior knowledge of SNR during inference and performs with higher efficiency. Simulation results demonstrated that the proposed method can reduce latency by about 90\% and improve accuracy by at most 2.65 dB than the baseline diffusion method, indicating its suitability for the practical deployment in future wireless communication systems.


\bibliographystyle{IEEEtran}
\bibliography{string}
\vspace{12pt}
\end{document}